\documentclass[a4paper,10pt]{article}

\usepackage{amsmath}
\usepackage{amssymb}
\usepackage{graphics}
\usepackage{rotating}
\usepackage{cite}
\usepackage{color}
\title{A VERY BRIEF NOTE ON SOME COMMUTATIVE ALGEBRAIC PROPERTIES OF A DIRAC-FUETER MODIFIED EQUATION.}
\author{Daniel Alay\'{o}n-Solarz,\underline{(solarz@ime.unicamp.br)} }

\begin{document}

\date{{\em December 02, 2004}}
\maketitle

\begin{abstract}We call attention to the unusual properties that the 4 dimensional solutions for a modified Fueter-Dirac equations satisfy: In a coordinate-free, constant-free and strictly mathematical way it is possible to show that all the solutions for a modified Fueter-Dirac Equation,
which are radial symmetric when restricted to the 3-space, have a nice algebraic structure. Locally, these solutions behave
naturally in a algebraic way determined by a certain commutative ring related to the quaternions with non-invertibles. This ring has a nontrivial localization. By using left and right versions of the operator we obtain quirality.
\end{abstract}

\maketitle

%%%%%%%%%%%%%%%%%%%%%%%%%%%%%%%%%%%%%%%%%%%%%%%%%%%%%%%%%%%%%%%%%%%%%%%%%%%%%%%
%                                SECTION I
%%%%%%%%%%%%%%%%%%%%%%%%%%%%%%%%%%%%%%%%%%%%%%%%%%%%%%%%%%%%%%%%%%%%%%%%%%%%%%%

We consider the following differential operator of order 1:

\begin{equation}
\dfrac{\partial f}{\partial t} + i \dfrac{\partial f}{\partial x} + j
\dfrac{\partial f}{\partial y} + k \dfrac{\partial f}{\partial z} = \dfrac{-2v}{r}
\end{equation}
restricted to the quaternionic functions, that satisfy:
\begin{equation}
f(p) = u(p) + \iota_{p}v(p)
\end{equation}
Where $u$ and $v$ are real functions of a quaternionic variable.
These functions satify the following properties:

\textbf{P1. (4D NON-OBSERVABILITY)} A solution that is quaternion
inversible almost everywhere has a singularity of codimension at most
2, a singularity the set where the function is not a local
diphemorphism.

\textbf{P1. (3D OBSERVABILITY)} A solution whose singularity has
codimension 2 will separate the imaginary space in $n$ connected
components.
\textbf{P1.1. (TOPOLOGICAL CHARACTERIZATION)} The number $n$ above
defined is a topological invariant.

\textbf{P2. (3D RADIAL SYMMETRY)}: A solution is radially symmetric
when restricted to the imaginary quaternionic space, isomorph to
$R^{3}$.

\textbf{P3. (4D INVARIANCE)}: Solutions are invariant under
quaternionic automorphisms.

\textbf{P4. (ALGEBRAIC QUATERNIONIC CLOSEDNESS)} Two solutions commute as quaternions. The sum and quaternionic product of two solutions is again a solution. The if the solution is locally algebraic inversible at some point then the algebraic inverse function is a solution.

\textbf{P5. (SEPARATION OF GLOBAL FUNCTIONS)} A solution defined on
the whole quaternions without the origin is either a local
diphemorphism almost everywhere or its totally degenerate in this
sense.

\textbf{P6. (QUIRAL CAUCHY-RIEMANN EQUATIONS)} It is always possible
to find a local parametrization based on 4 real variables variables,
denoted as $t$ ,$r$ ,$\alpha$, $\beta$ such that the function will
satisfy the following 4 differential equations:
\begin{eqnarray}
\dfrac{\partial u}{\partial t} - \dfrac{\partial v}{\partial r} = 0
\\
\dfrac{\partial u}{\partial r} + \dfrac{\partial v}{\partial t} = 0
\\
\dfrac{\partial v}{\partial \alpha}(\sin\beta)^{-2}+\dfrac{\partial u}{\partial \beta} = 0
\\
\dfrac{\partial u}{\partial \alpha}(\sin\beta)^{-2}-\dfrac{\partial v}{\partial \beta}= 0
\end{eqnarray}
and the following subproperty:

\textbf{P4*} If $\dfrac{\partial f}{\partial t} = \dfrac{\partial f}{\partial
r} = 0$ then $f$ is not locally invertible.

\textbf{P4**} If $\dfrac{\partial f}{\partial \alpha} = \dfrac{\partial
f}{\partial \beta} = 0$ then $f$ is locally invertible unless
constants.

Thus the solutions are a commutative ring with quaternionic product and topological non-inversible solutions, where the commutative algebra machinery of localization seems natural to apply as the localization is not trivial. Quirality is obtained once we consider left and right versions of this operator.

%%%%%%%%%%%%%%%%%%%%%%%%%%%%%%%%%%%%%%%%%%%%%%%%%%%%%%%%%%%%%%%%%%%%%%%%%%%%%%
\end{document}